\documentclass[aps,prb,twocolumn,preprintnumbers,amsmath,amssymb,superscriptaddress]{revtex4}%

\usepackage{graphicx}
\usepackage{dcolumn}
\usepackage{amsmath}
\usepackage{color}
\usepackage{hyperref}
\usepackage{bbold}
\usepackage{multirow}

\newcommand{\RN}[1]{\textup{\uppercase\expandafter{\romannumeral#1}}}%

\begin{document}

\title{Oxygen-isotope effect on density wave transitions in La$_3$Ni$_2$O$_{7}$}

\author{Rustem Khasanov}
 \email{rustem.khasanov@psi.ch}
 \affiliation{PSI Center for Neutron and Muon Sciences CNM, 5232 Villigen PSI, Switzerland}

\author{Vahid Sazgari}
 \affiliation{PSI Center for Neutron and Muon Sciences CNM, 5232 Villigen PSI, Switzerland}

\author{Igor Plokhikh}
 \affiliation{PSI Center for Neutron and Muon Sciences CNM, 5232 Villigen PSI, Switzerland}
 \affiliation{TU Dortmund University, Department of Physics, Dortmund, 44227, Germany}

\author{Lifen Shi}
 \affiliation{Max Planck Institute for Chemical Physics of Solids, N\"{o}thnitzer Strasse 40, 01187 Dresden, Germany}

\author{KeYuan Ma}
 \affiliation{Max Planck Institute for Chemical Physics of Solids, N\"{o}thnitzer Strasse 40, 01187 Dresden, Germany}

\author{Marisa Medarde}
 \affiliation{PSI Center for Neutron and Muon Sciences CNM, 5232 Villigen PSI, Switzerland}

\author{Ekaterina Pomjakushina}
 \affiliation{PSI Center for Neutron and Muon Sciences CNM, 5232 Villigen PSI, Switzerland}

\author{Tomasz Klimczuk}
 \affiliation{Faculty of Applied Physics and Mathematics, Gda\'{n}sk University of Technology, Narutowicza 11/12, Gdansk, 80-233 Poland}
 \affiliation{Advanced Materials Center, Gda\'{n}sk University of Technology, Narutowicza 11/12, Gdansk, 80-233 Poland}

\author{Thomas J. Hicken}
 \affiliation{PSI Center for Neutron and Muon Sciences CNM, 5232 Villigen PSI, Switzerland}

\author{Hubertus Luetkens}
 \affiliation{PSI Center for Neutron and Muon Sciences CNM, 5232 Villigen PSI, Switzerland}

\author{Christof W. Schneieder}
 \affiliation{PSI Center for Neutron and Muon Sciences CNM, 5232 Villigen PSI, Switzerland}

\author{Zurab Guguchia}
 \affiliation{PSI Center for Neutron and Muon Sciences CNM, 5232 Villigen PSI, Switzerland}

\author{Sergey Medvedev}
 \affiliation{Max Planck Institute for Chemical Physics of Solids, N\"{o}thnitzer Strasse 40, 01187 Dresden, Germany}

\author{Dariusz J. Gawryluk}
 \affiliation{PSI Center for Neutron and Muon Sciences CNM, 5232 Villigen PSI, Switzerland}

\begin{abstract}
The isotope effect is a powerful probe of electron–phonon interactions in solid-state systems, offering key insights into how atomic mass influences emergent quantum states.
Here, the impact of oxygen isotope substitution ($^{16}{\rm O}\rightarrow \; ^{18}{\rm O}$) on charge- and spin-density wave (CDW and SDW) transitions in the double-layer Ruddlesden–Popper nickelate La$_3$Ni$_2$O$_7$ is investigated.
A clear isotope effect is observed in the CDW transition: the transition temperature ($T_{\rm CDW}$) increases upon $^{18}$O substitution.
In contrast, the SDW transition temperature remains unaffected within experimental uncertainty.
These findings point to a strong involvement of lattice vibrations in the formation of charge order, while spin order appears to be predominantly of electronic origin.
The results suggest that electron–phonon coupling, manifested through the CDW response to isotope substitution, may be relevant to the superconducting pairing mechanism in Ruddlesden–Popper nickelates.
\end{abstract}
\maketitle

\noindent \underline{{\it Introduction.}}
The isotope effect was recognized as a fundamental tool in condensed matter physics for probing the intricate relationship between lattice vibrations and electronic properties. By replacing atoms with their isotopic counterparts, the role of phonons in various electronic and magnetic phase transitions can be assessed. This approach is particularly effective in differentiating phonon-driven phenomena from purely electronic interactions in complex materials.

Historically, the isotope effect was first observed in conventional superconductors, where it played a pivotal role in confirming the phonon-mediated pairing mechanism described by Bardeen-Cooper-Schrieffer (BCS) theory.\cite{frohlich_PR_1950, BCS_theory_1957} In these materials, the superconducting transition temperature ($T_c$) follows the empirical relation $T_c \propto M^{-\alpha}$, where $M$ represents the isotopic mass and the isotope exponent $\alpha \approx 0.5$ for an ideal phonon-mediated mechanism.\cite{Maxwell_PR_1950, Reynolds_PR_1951, Olsen_Cryogenics_1963, Shaw_Physrev_1961, Mathias_PR_1963, Bucher_PhysRevA_1967, Fassnacht-PRL_1966, Budko_PRL_2001, Hinks_Nature_2001, Hein _PhysRev_1963, Geballe_PRL_1961, Finnemore_PRL_1962, Gibson_PhysRev_1966, Gebale_IBM-Res_1962, Fowler_PRL_1967} This dependency arises because phonons mediate Cooper pair formation, and their characteristic frequencies scale as $M^{-1/2}$. However, significant deviations from this standard isotope effect was reported in strongly correlated electron systems, where additional interactions beyond conventional electron-phonon coupling are believed to contribute.\cite{Geballe_PRL_1961, Finnemore_PRL_1962, Gibson_PhysRev_1966,Fowler_PRL_1967,  Stritzker_ZPhys_1972} In unconventional superconductors, the isotope effect varies significantly, suggesting a complex interplay between lattice vibrations and superconducting carriers.\cite{Batlogg_PRL_1987, Faltens_PRL_1987, Batlogg_PRL2_1987, Bourne_PRL_1987, Franck_PRB_1991, Babushkina_PhysicaC_1991, Zhao_Nature_1997, Zech_Nature_1994, Zech_PhysicaB_1996, Khasanov_PRB_SSOIE_2003, Tallon_PRL_2005, Keller_MaterialsToday_2008, Khasanov_PRL_LEM-OIE_2004, Mao_PRB_2001, Khasanov_PRB_OIE-general_2006, Khasanov_PRL_YPa123_2008, Rischau_PRR_2022, Schlueter_PhysicaC_2001, Liu_nature_2009, Shirage_PRL_2009, Khasanov_NJP_2010, Khasanov_PRB_IE-FeSe_2010}

Beyond superconductivity, isotope substitution was shown to influence the charge-density wave (CDW) and spin-density wave (SDW) transition temperatures in various materials. \cite{Khasanov_PRL_YPa123_2008, Guguchia_PRL_2014, Medarde_PRL_1998, Luetkens_JMMM_2007, Amit_AdvCondMat_2011, Shengelaya_PRL_1999, Zhao_PRB_1994, Lanzara_JPCM_1999, Guguchia_PRB_2015, Bendele_PRB_2017} The formation of charge and spin-density waves arises due to collective electronic instabilities, often accompanied by lattice distortions, making these phases highly sensitive to phononic contributions. Variations in the density wave transition temperatures ($T_{\text{CDW}}$ and $T_{\text{SDW}}$) upon isotope substitution provide insight into the relative importance of electron-phonon interactions versus purely electronic effects in stabilizing these ordered states.

The recently discovered Ruddlesden-Popper (RP) nickelate superconductors exhibit a rich variety of electronic phases, with some of them potentially serving as precursors to superconductivity. In double- and tri-layer RP nickelates, charge density wave (CDW) and spin density wave (SDW) orders are observed at ambient pressure, while superconductivity emerges only under high-pressure conditions.\cite{Sun_Nature_2023, Zhang_NatPhys_2024, Wang_Nature_2024, Liu_NatCom_2024, Zhang_JMST_2024, Li_SciBull_2024, Li_ChinPhysLet_2024, Zhou_arxiv_2023, Zhou_arxiv_2024, Sakakibara_PRB_2024, Wang_ChinPhysLett_2024, Pei_arxiv_2024, Wang_PRX_2024, Li_arxiv_2025, Puphal_PRL_2024, Zhu_Nature_2024, Zhang_PRX_2025, Khasanov_NatPhys_La327_2025}
Since superconductivity in these materials arises from states dominated by CDW and SDW correlations, investigating the isotope effect on both the precursor density wave orders and the superconducting phase offers valuable insights into the underlying pairing mechanism.

This work presents an investigation of the oxygen isotope effect (OIE) on the density-wave transition temperatures in the double-layer RP nickelate La$_3$Ni$_2$O$_7$.
Our experiments reveal that the charge-density-wave transition is sensitive to isotope substitution, resulting in a higher transition temperature $T_{\rm CDW}$ for the sample containing the heavier oxygen isotope. The corresponding isotope shift, $\Delta T_{\rm CDW} = T_{\rm CDW}^{18} - T_{\rm CDW}^{16}$, is approximately $2.3$~K (throughout this study, superscripts $^{16}$ and $^{18}$ are used to indicate substitution with the corresponding oxygen isotope). In contrast, the isotope shift of the spin-density-wave  transition temperature is negligible within experimental uncertainty, with $\Delta T_{\rm SDW} = T_{\rm SDW}^{18} - T_{\rm SDW}^{16} = -0.08(9)$~K.
These findings indicate that lattice vibrations play a significant role in the formation or stabilization of the CDW state, whereas the SDW transition appears to be governed predominantly by electronic interactions. The contrasting isotope responses of the CDW and SDW transitions highlight the different microscopic origins of these ordered states in La$_3$Ni$_2$O$_7$ and provide useful constraints for theoretical models of density-wave formation in Ruddlesden–Popper nickelates.


\noindent \underline{{\it Sample preparation and experimental techniques.}}
The sample preparation procedure, as well as the experimental techniques used in the present study -- namely x-ray diffraction, thermogravimetry, Raman spectroscopy, muon-spin rotation/relaxation ($\mu$SR), and resistivity -- are described in the Supplemental Material part.\cite{Supplemental_Material}

A schematic of the oxygen-isotope substitution apparatus is shown in Fig.~S1(a) in the Supplemental Material, Ref.~\onlinecite{Supplemental_Material}, and described in detail in Refs.~\onlinecite{Conder_MatScIng_2001, Conder_PhysicaC_2023}. The $^{18}$O content in the La$_3$Ni$_2\,^{18}$O$_{7\pm\delta}$ sample was determined to be 82(2)\% using mass spectrometry analysis of the $^{18}$O$_2$ gas line.

X-ray diffraction experiments confirmed that the crystal structures of the $^{16}$O- and $^{18}$O-substituted sample pair remain unchanged. The results, including the symmetry group and lattice parameters ($a$, $b$, and $c$), are presented in Fig.~S1(b) and Table~S1 of the Supplemental Material.\cite{Supplemental_Material} Isotope substitution was found to have a negligible effect on all lattice parameters of La$_3$Ni$_2$O$_7$.

The oxygen content in the $^{16}$O- and $^{18}$O-substituted La$_3$Ni$_2$O$_{7\pm\delta}$ samples was determined using thermogravimetric analysis [Fig.~S1(c) in the Supplemental Material, Ref.~\onlinecite{Supplemental_Material}]. The samples were found to be nearly oxygen-stoichiometric, with oxygen contents ($7 \pm \delta$) determined to be 6.99(1) and 7.04(2) for La$_3$Ni$_2\,^{16}$O$_{7\pm\delta}$ and La$_3$Ni$_2\,^{18}$O$_{7\pm\delta}$, respectively. Hereafter, these samples are referred to as La$_3$Ni$_2\,^{16}$O$_7$ and La$_3$Ni$_2\,^{18}$O$_7$.

\begin{figure}[htb]
\includegraphics[width=0.9\linewidth]{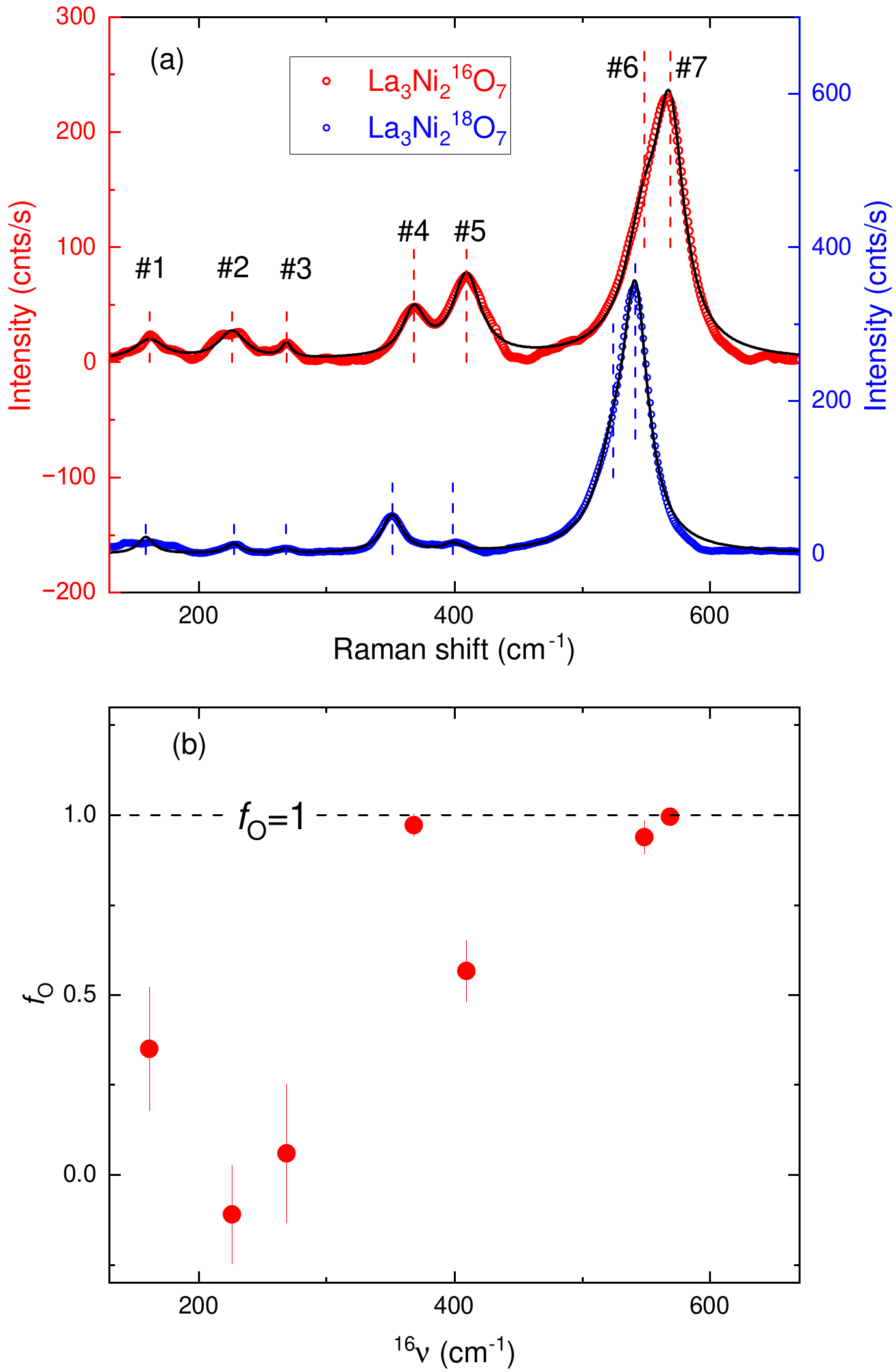}
\caption{
(a) Room-temperature Raman spectra of oxygen-isotope substituted La$_3$Ni$_2$O$_7$ samples. Solid lines represent fits using
seven Lorentzian functions. Dashed lines indicate the peak positions.
(b) The oxygen participation parameter $f_{\rm O}$ as a function of the Raman wavenumber $^{16}\nu$.
}
 \label{fig:Raman_experiments}
\end{figure}

\noindent \underline{{\it Oxygen-isotope effect on the phonon frequencies.}}
Room-temperature Raman spectroscopy was employed to confirm both the effectiveness and spatial uniformity of $^{18}$O isotope substitution, taking advantage of the technique’s sensitivity to mass-dependent shifts in phonon frequencies.
The measured spectra, shown in Fig.~\ref{fig:Raman_experiments}~(a), were fitted using a sum of seven Lorentzian-shaped peaks. The extracted phonon frequencies for the $^{16}$O- and $^{18}$O-substituted La$_3$Ni$_2$O$_7$ samples, denoted as $\nu^{16}$ and $\nu^{18}$, respectively, are listed in Table~S2 in the Supplemental Material.\cite{Supplemental_Material}

The coefficient representing the involvement of oxygen in each phonon mode, $f_{\rm O}$, was calculated following Refs.~\onlinecite{Cardona_RMP_2005, Menedez_Philmag_1994, Zhang_PRB_1997}:
\begin{equation}
\nu^{18}(x)/\nu^{16} = \left[\sqrt{M^{16}/M^{18}(x)}\right]^{f_{\rm O}},
\label{eq:f_O}
\end{equation}
where $x$ is the $^{18}$O content in the isotope-substituted sample (82\% in this study), $M^{16} = 16$ is the atomic mass of $^{16}$O, and $M^{18}(x) = (1 - x)\cdot M^{16} + x\cdot M^{18}$ is the average oxygen mass in the $^{18}$O-substituted sample. This relationship assumes a harmonic oscillator model, where mode softening upon isotope substitution reflects the oxygen contribution to each vibration.

Based on the extracted $f_{\rm O}$ values shown in Fig.~\ref{fig:Raman_experiments}~(b) and listed in Table~S2 of the Supplemental Material,\cite{Supplemental_Material} the following conclusions can be drawn:\\
(i) Three phonon modes observed at $\nu^{16}$ ($\nu^{18}$) = 368 (351), 549 (524), and 568 (542)~cm$^{-1}$ exhibit $f_{\rm O}$ values close to 1.0, indicating that these modes are dominated by oxygen vibrations.\\
(ii) The remaining modes at 161 (159), 226 (227), 269 (268), and 409 (398)~cm$^{-1}$ show significantly smaller -- or even negligible, within experimental uncertainty -- values of $f_{\rm O}$, implying that oxygen is only weakly involved or not involved at all in these lattice vibrations. These modes are likely associated with heavier ions, such as Ni and La.

\begin{figure}[htb]
\includegraphics[width=0.9\linewidth]{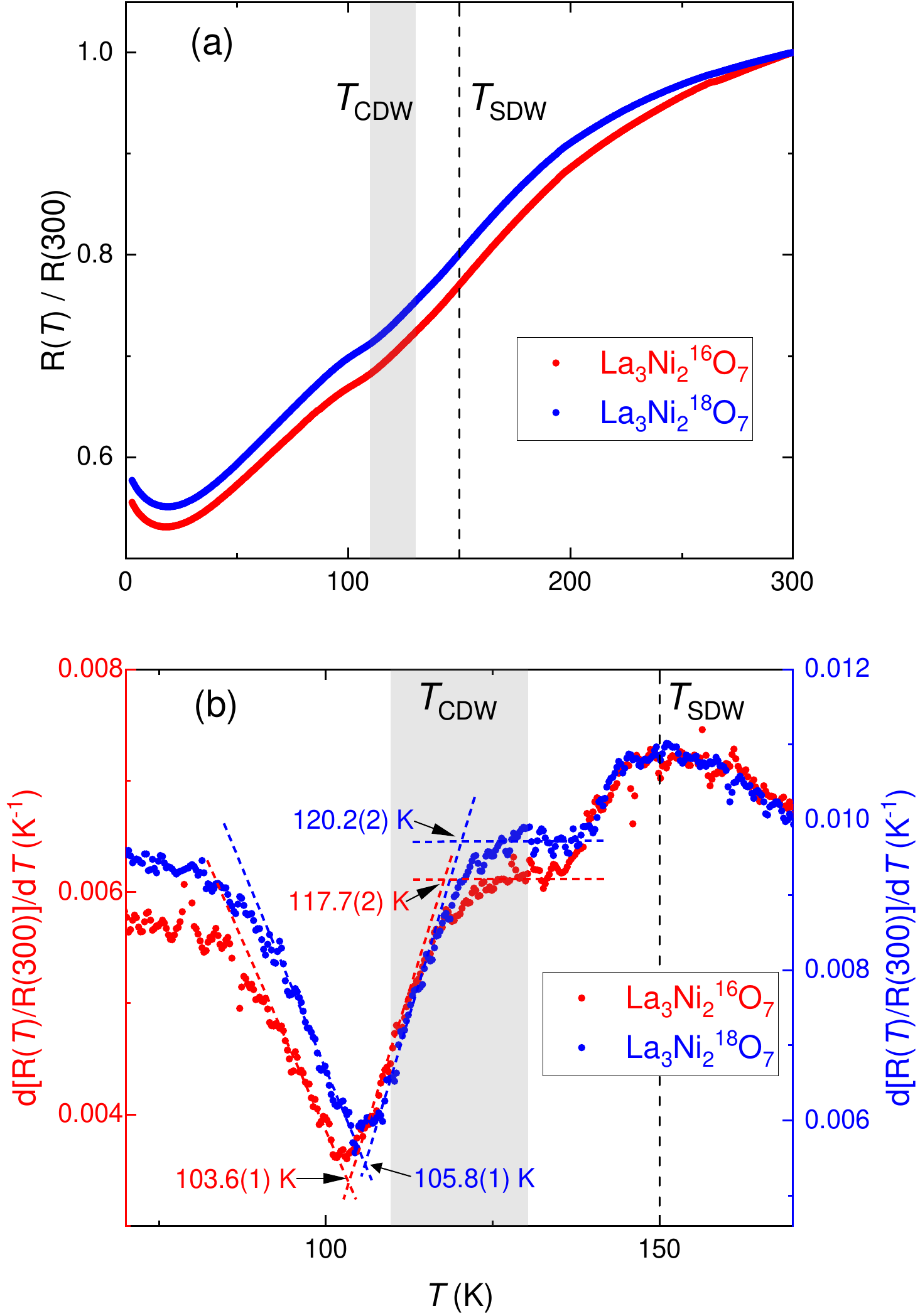}
\caption{
(a) Raw resistivity curves of the $^{16}$O/$^{18}$O-substituted La$_3$Ni$_2$O$_7$ samples.
(b) Extended view of the first derivatives of the resistivity data in the vicinity of the CDW transitions. The red and blue numbers correspond to the CDW transition temperatures of La$_3$Ni$_2\,^{16}$O$_7$ and La$_3$Ni$_2\,^{18}$O$_7$, determined using different criteria (see text for details).
}
 \label{fig:Resistivity-experiments}
\end{figure}

\noindent\underline{{\it Oxygen isotope effect on the CDW transition.}}
The oxygen isotope effect on the CDW transition temperature was investigated via resistivity measurements. Figure~\ref{fig:Resistivity-experiments}~(a) shows the temperature dependence of resistivity normalized to its 300~K value, $R(T)/R(300)$. The raw resistivity curves exhibit weak anomalies near $T \sim 120$~K. A more pronounced structure emerges in the first derivative of the resistivity data, shown in Fig.~\ref{fig:Resistivity-experiments}~(b).

The anomalies observed in both the $R(T)/R(300)$ and ${\rm d}[R(T)/R(300)]/{\rm d}T$ curves may be attributed to charge- and spin-density wave transitions [dashed lines and gray stripe in Figs.~\ref{fig:Resistivity-experiments}~(a) and (b)], as previously reported in the literature. The SDW transition at $T_{\rm SDW} \simeq 150$~K was identified in earlier $\mu$SR, NMR, NQR, and resonant x-ray studies.\cite{Chen_PRL_2024, Khasanov_NatPhys_La327_2025, Chen_arxiv_2024, Dan_SciBull_2025, Kakoi_JPSJ_2024, Chen_NatPhys_2024, Ren_CommPhys_2025, Gupta_NatCom_2025, Luo_CinPhysLett_2025} Although the CDW in La$_3$Ni$_2$O$_7$ has not yet been directly confirmed by scattering techniques, anomalies in resistivity and specific heat at $110 \lesssim T \lesssim 130$~K have been associated with the onset of CDW order.\cite{Liu_NatCom_2024, Wang_ChinPhysLett_2024, Wang_InorgChem_2024, Wu_PRB_2001, Liu_SciChina_2023, Seo_InorgChem_1996} Accordingly, an average CDW transition temperature of $T_{\rm CDW} = 120$~K is adopted for La$_3$Ni$_2$O$_7$. It is also worth noting that recent optical pump-probe spectroscopy,\cite{Meng_NatCom_2024} x-ray absorption near-edge spectroscopy,\cite{LiMintago_arxiv_2025} and NQR\cite{Luo_CinPhysLett_2025} measurements have provided further indications of CDW order below 150~K in La$_3$Ni$_2$O$_7$.

The isotope-induced shift of the CDW transition temperature was estimated using two approaches:
(i) by determining the crossing point between linear fits to the derivative curves within the `slope' and `plateau' regions, which yielded $T_{\rm CDW}^{16} = 117.7(2)$~K and $T_{\rm CDW}^{18} = 120.2(2)$~K; and
(ii) by identifying the crossing points of linear fits around the local minimum, giving $T_{\rm CDW}^{16} = 103.6(1)$~K and $T_{\rm CDW}^{18} = 105.8(1)~K$.
Both approaches yield a consistent isotope shift of $\Delta T_{\rm CDW} = T_{\rm CDW}^{18} - T_{\rm CDW}^{16} \simeq 2.4(2)$~K.

It should also be noted that the derivative curves exhibit a broad hump around 150~K, which may be associated with the SDW transition, as reported in Refs.~\onlinecite{Chen_PRL_2024, Khasanov_NatPhys_La327_2025, Chen_arxiv_2024, Dan_SciBull_2025, Kakoi_JPSJ_2024, Chen_NatPhys_2024, Ren_CommPhys_2025, Gupta_NatCom_2025, Luo_CinPhysLett_2025}. However, this feature is too broad to allow for a reliable determination of the OIE on $T_{\rm SDW}$ from resistivity data alone.
Instead, the isotope effect on $T_{\rm SDW}$ is evaluated based on $\mu$SR measurements, presented below.

\noindent \underline{{\it Oxygen isotope effect on the SDW transition.}}
The OIE on the SDW transition temperature was investigated using $\mu$SR. Experiments were conducted in the weak-transverse field (WTF) mode with a magnetic field $B_{\rm WTF}=5$~mT applied perpendicular to the initial muon-spin polarization. The data analysis procedure, along with several WTF data sets, are provided in the Supplemental Material.\cite{Supplemental_Material}

\begin{figure}[htb]
\includegraphics[width=0.85\linewidth]{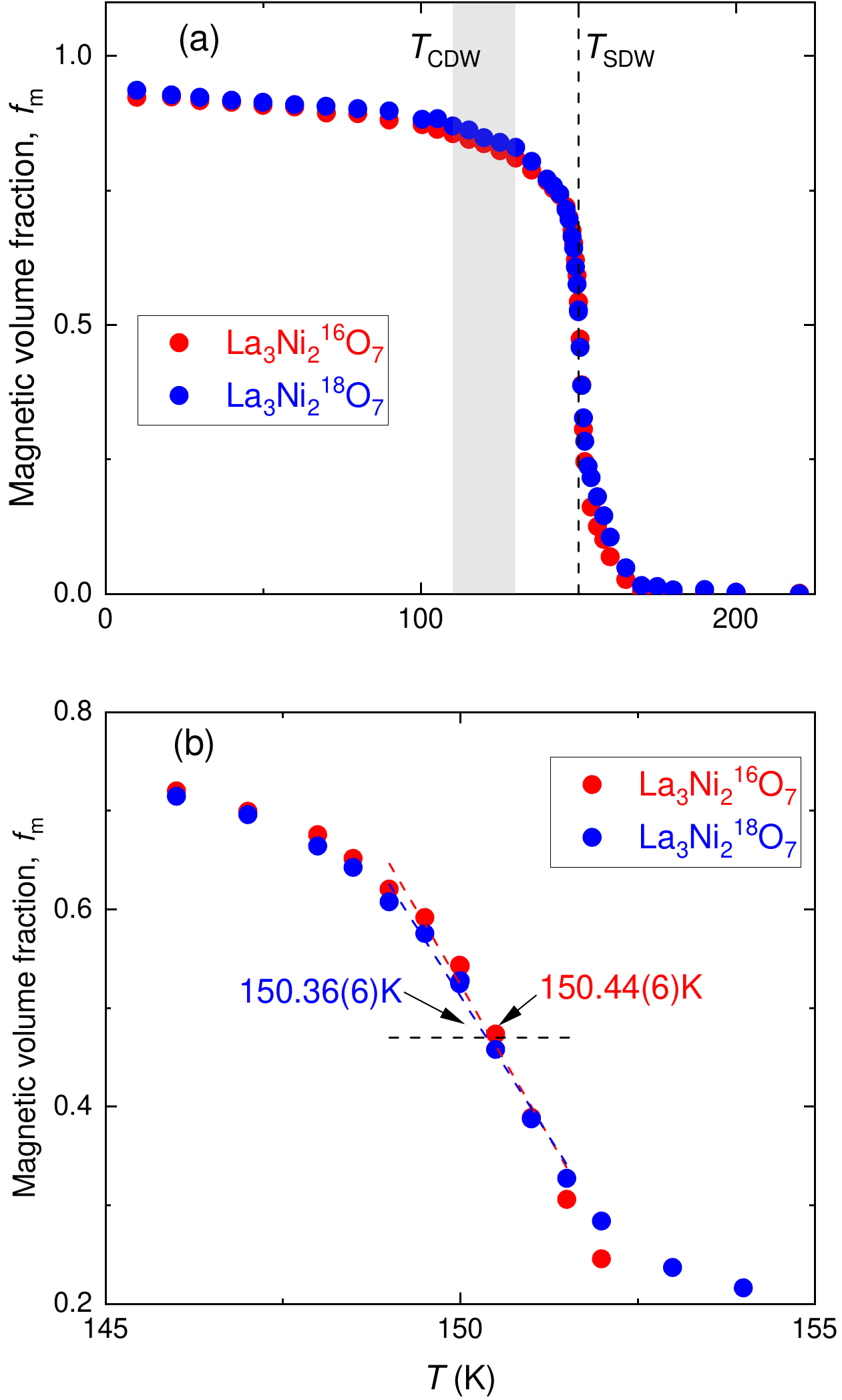}
\caption{
(a) Temperature dependence of the magnetic volume fraction $f_{\rm m}$ for $^{16}$O- and $^{18}$O-substituted La$_3$Ni$_2$O$_7$ and La$_4$Ni$_3$O$_{10}$ samples.
(b) Extended view of the $f_{\rm m}(T)$ curves in the vicinity of the SDW transitions.
The vertical dashed lines and grey stripe indicate the CDW and SDW transition temperatures ($T_{\rm CDW}$ and $T_{\rm SDW}$), as reported in the literature.\cite{Chen_PRL_2024, Khasanov_NatPhys_La327_2025, Chen_arxiv_2024, Zhang_NatCom_2020, Khasanov_arxiv_La4310_2025, Liu_NatCom_2024, Wang_ChinPhysLett_2024, Wang_InorgChem_2024, Wu_PRB_2001, Liu_SciChina_2023, Seo_InorgChem_1996}}
 \label{fig:WTF-experiments}
\end{figure}

The temperature evolution of the magnetic volume fractions obtained from the fits to WTF-$\mu$SR data is presented in Fig.~\ref{fig:WTF-experiments}~(a). The dashed lines and grey stripe indicate the CDW and SDW transition temperatures, as reported in the literature.\cite{Chen_PRL_2024, Khasanov_NatPhys_La327_2025, Chen_arxiv_2024, Zhang_NatCom_2020, Khasanov_arxiv_La4310_2025, Liu_NatCom_2024, Wang_ChinPhysLett_2024, Wang_InorgChem_2024, Wu_PRB_2001, Liu_SciChina_2023, Seo_InorgChem_1996}  The WTF data clearly capture the SDW transition.
It should be noted that the maximum value of the magnetic volume fraction does not reach 100\%, which is attributed to approximately 5-7\% of the muons missing the sample and stopping in the sample holder and/or the cryostat walls.

Figure~\ref{fig:WTF-experiments}~(b) shows an expanded view of the $f_{\rm m}(T)$ curves in the vicinity of $T_{\rm SDW}$. The SDW transition temperatures were estimated from the intersection of linearly extrapolated $f_{\rm m}(T)$ curves near $T_{\rm SDW}$ with the reference line $f_{\rm m}(T) = 0.5 \cdot f_{\rm m}(10~\text{K})$ [dashed lines in Fig.~\ref{fig:WTF-experiments}~(b)]. The fits yield $T_{\rm SDW}^{16} = 150.44(6)$~K and $T_{\rm SDW}^{18} = 150.36(6)$~K, resulting in an isotope shift of $\Delta T_{\rm SDW} = T_{\rm SDW}^{18} - T_{\rm SDW}^{16} = -0.08(9)$~K.

\begin{table*}[htb]
\caption{Results of oxygen isotope effect measurements of La$_3$Ni$_2$O$_7$. $\Delta T_{tr}=T_{tr}^{18}-T_{tr}^{16}$ ($T_{tr}= T_{\rm SDW}$ or $T_{\rm CDW}$). The oxygen isotope exponent $\alpha$ is calculated by using equation  $\alpha=-{\rm d}\ln T_{tr}/{\rm d}\ln M$ and is corrected for uncomplete isotope exchange $\simeq 82$\%.  }
\begin{tabular}{c|c|c|c|c|c}
    & La$_3$Ni$_2^{16}$O$_7$& La$_3$Ni$_2^{18}$O$_7$&$\Delta T_{tr}$&$\alpha$&Technique\\
  \hline
$T_{\rm SDW}$ & 150.44(6)~K& 150.36(6)~K& -0.08(9)~K& 0.005(6)& WTF-$\mu$SR\\
\multirow{2}{*}{$T_{\rm CDW}$}& 103.6(1)~K\footnotemark[1]&105.8(1)~K\footnotemark[1]&2.2(1)~K&-0.21(1)&Resistivity\\ &117.7(2)~K\footnotemark[2]&120.2(2)~K\footnotemark[2]&2.5(4)~K&-0.21(3)&Resistivity\\
\end{tabular}
\footnotetext[1]{From local minima at ${\rm d}[R(T)/R(300)]/{\rm d}T$ [Fig.~\ref{fig:Resistivity-experiments}~(b)]}
\footnotetext[2]{From linear slopes of ${\rm d}[R(T)/R(300)]/{\rm d}T$ [Fig.~\ref{fig:Resistivity-experiments}~(b)]}
\label{table:OIE-results}
\end{table*}

\noindent \underline{{\it Discussion and Conclusions.}}
Table~\ref{table:OIE-results} summarizes the results of the oxygen isotope effect measurements on the charge- and spin-density wave transition temperatures in the double-layer Ruddlesden--Popper nickelate La$_3$Ni$_2$O$_7$.
The OIE results reported in the present study for La$_3$Ni$_2$O$_7$, together with those in Ref.~\onlinecite{Khasanov_arxiv_La4310_2025} for the tri-layer RP nickelate La$_4$Ni$_3$O$_{10}$, can be directly compared to isotope effects investigated in cuprate high-temperature superconductors (HTSs), where such studies are among the most extensive within correlated electron systems.

In undoped cuprates, the isotope effect on the N\'{e}el temperature $T_N$ is found to be negligible,\cite{Khasanov_PRL_YPa123_2008, Zhao_PRB_1994, Amit_AdvCondMat_2011}
consistent with the observations for La$_3$Ni$_2$O$_7$, where the SDW transition -- preceding the CDW order -- shows no measurable isotope shift. This suggests that magnetism, in the absence of strong charge fluctuations, is primarily governed by electronic interactions and is only weakly influenced by phonons.

In contrast, underdoped cuprates, where charge and spin orders coexist and are strongly correlated, show a different behavior. In stripe-ordered cuprates such as La$_{2-x}$Ba$_x$CuO$_4$, the emergence of static charge order is accompanied by a spin-stripe phase that strongly suppresses superconductivity.\cite{Moodenbaugh_PRB_1988, Sears_PRB_2023, Hu_PRB_2025, Abbamonte_NatPhys_2005, Tranquada_Nature_1995, Fujita_PRB_2004, Hucker_PRB_2011, Coboz_PRL_2014}
Notably, isotope effect studies in these systems show that both the CDW and SDW transition temperatures increase upon $^{18}$O substitution, with shifts of the same sign and comparable magnitude.\cite{Guguchia_PRL_2014, Guguchia_PRB_2015}

This behavior is strikingly similar to the findings in La$_4$Ni$_3$O$_{10}$,\cite{Khasanov_arxiv_La4310_2025} where CDW and SDW orders are intertwined, share the same transition temperature ($T_{\rm CDW} \simeq T_{\rm SDW}$), and exhibit identical isotope shifts ($\Delta T_{\rm CDW} = \Delta T_{\rm SDW}$).
These parallels suggest a shared underlying mechanism, likely involving strong electron--phonon coupling that affects both spin and charge degrees of freedom when the two orders are strongly entangled.

Despite these parallels, there are notable differences between cuprates and nickelates. In cuprate HTSs, increasing hole doping suppresses the SDW order, eventually leading to the emergence of superconductivity,\cite{Damascelli_RMP_2003, Taillefer_AnnRev_2010, Keimer_Nature_2015} while in La$_3$Ni$_2$O$_7$, applying pressure does not immediately suppress SDW order.\cite{Khasanov_NatPhys_La327_2025}
Moreover, in La$_4$Ni$_3$O$_{10}$, the SDW order does not appear to be the primary instability, but rather a consequence of CDW formation.\cite{Khasanov_arxiv_La4310_2025}

In kagome materials, such as CsV$_3$Sb$_5$, KV$_3$Sb$_5$, and RbV$_3$Sb$_5$, suppression of CDW order through pressure or chemical tuning was found to enhance superconductivity.\cite{Chen_PRL_2021, Yu_NatPhys_2021, Zheng_Nature_2022, Guguchia_NatCom_2023, Yang_SciBull_2022} A similar trend may be expected in nickelates, where superconductivity emerges under high pressure and is associated with the suppression of charge order. This raises the possibility that, as in kagome superconductors, CDW order plays the dominant role in competing with superconductivity in RP nickelates, rather than SDW order.


To conclude, our findings reveal that the CDW and SDW transitions in La$_3$Ni$_2$O$_7$ exhibit markedly different sensitivities to oxygen-isotope substitution. The clear isotope dependence of the CDW transition highlights a strong coupling between charge order and lattice vibrations, while the SDW transition shows no measurable isotope shift, indicating a predominantly electronic origin. This distinct isotope response offers valuable insight into the role of electron--phonon interactions and electronic correlations in layered nickelates, and demonstrates that isotope substitution is an effective tool for disentangling competing ordered states in Ruddlesden--Popper systems.

\noindent \underline{{\it Acknowledgments}} Z.G. acknowledges support from the Swiss National Science Foundation (SNSF) through SNSF Starting Grant (No. TMSGI2${\_}$211750). I.P. acknowledges financial support from Paul Scherrer Institute research grant No. 2021\_0134. D.J.G. acknowledges support from the Swiss National Science Foundation (SNSF) through Grant No. 200021E\_238113.

\end{document}